# Novel algorithms and high-performance cloud computing enable efficient fully quantum mechanical protein-ligand scoring

Narbe Mardirossian[1*], Yuhang Wang[2], David A. Pearlman[2], Garnet Kin-Lic Chan[3*], and Toru Shiozaki[2*]

[1]AMGEN Research, One Amgen Center Drive, Thousand Oaks, CA 91320

[2]Quantum Simulation Technologies, Inc., Cambridge, MA 02139

[3]Division of Chemistry and Chemical Engineering, California Institute of Technology, Pasadena, CA 91125

*nmardiro@amgen.com, gkc1000@gmail.com, shiozaki@qsimulate.com

## Abstract

Ranking the binding of small molecules to protein receptors through physics-based computation remains challenging. Though inroads have been made using free energy methods, these fail when the underlying classical mechanical force fields are insufficient. In principle, a more accurate approach is provided by quantum mechanical density functional theory (DFT) scoring, but even with approximations, this has yet to become practical on drug discovery-relevant timescales and resources. Here, we describe how to overcome this barrier using algorithms for DFT calculations that scale on widely available cloud architectures, enabling full density functional theory, without approximations, to be applied to protein-ligand complexes with approximately 2500 atoms in tens of minutes. Applying this to a realistic example of 22 ligands binding to MCL1 reveals that density functional scoring outperforms classical free energy perturbation theory for this system. This raises the possibility of broadly applying fully quantum mechanical scoring to real-world drug discovery pipelines.

## Introduction

Quantum mechanics (QM) is the fundamental physical theory that governs the microscopic world, from atoms to molecules to proteins. However, to date, it has only played a minor role in screening ligands in drug discovery processes in the pharmaceutical industry, mainly because the computational demands required for solving the quantum mechanical equations are far greater than those for the classical counterparts. Because of this, it has been asserted that full QM simulations of protein-ligand complexes, i.e., those that do not introduce additional approximations, are "challenging, if not impossible" on discovery-relevant timescales[1]. This communication aims to demonstrate that through new algorithms, robust software engineering, and the use of commodity cloud resources, we can now perform quantum mechanical density functional theory (DFT/QM) calculations of entire protein-ligand complexes with a turnaround time relevant to lead optimization, without introducing approximations to the quantum mechanics beyond those inherent to DFT/QM itself. We call this approach highly efficient distributed-memory QM (hedQM).

Computational methods in the drug discovery process can now be considered amongst the standard first-line approaches for moving from target identification to hits[2]. However, they have been less effective in later stages, i.e., going from hit-to-lead, where accurately ranking the binding affinities of multiple binders to a target protein is required. In the past decade, free energy calculation approaches, notably Free Energy Perturbation (FEP) and Thermodynamic Integration (TI)[3,4] have been demonstrated, for some systems, to be capable of providing predictions of suitable accuracy and with suitable turnaround such that they can be used to focus the type of costly bench synthesis that characterizes end-stage lead optimization[5,6]. Despite this, methods like FEP remain limited with respect to the systems for which they are reliably predictive. There are several reasons for this, but many are related to limitations in the classical force field that lies at the heart of these approaches. While decades of work have led to reasonably predictive force fields, it is well understood that the force fields in wide use still struggle with many common interactions, e.g. polarization, metal coordination, common substituents such as halogens, covalency, and others[7,8]. It is also difficult to reliably apply common free energy methods to systems where the formal charge is changing on the ligand or in the protein environment, or to a non-congeneric series of ligands binding to the same protein receptor[9].

One solution to these limitations of free energy calculations is to instead use accurate first-principles quantum mechanical methods, such as DFT/QM, for the characterization of ligand binding. DFT/QM obviates the need for either a force field form or atom-wise parameterization, can be used with arbitrary molecular moieties, and can deal natively with issues such as changes in valency, formal charge, and polarization. However, in the context of drug discovery time scales, full DFT/QM has not yet been practical to use in systems as large as protein-ligand complexes, and only more approximate, QM-derived approaches have been investigated in this context[1]. The two desirable but seemingly incompatible requirements are that (1) approximations should not suffer from similar limitations as force fields (or reintroduce errors on the scale of the few kcal/mol needed to discriminate between different ligands in realistic applications); (2) the computations should complete on the time scale of roughly an hour, to be relevant to real-world ligand optimization. While this has not been possible to date, the technical innovations we describe below now make it possible to carry out full DFT/QM calculations on protein-ligand complexes, without additional approximations, on time scales of a fraction of an hour.

For a practical demonstration, we select the problem of rank-ordering ligand binding to induced myeloid leukemia cell differentiation protein (MCL1). It is critical to note, at the outset, that rank ordering is fundamentally the use case of greatest significance in real-world small molecule drug discovery because it addresses the question of which compounds to synthesize. The MCL1 system was chosen for a variety of reasons: (1) crystal structures have been published with a representative bound ligand; (2) binding data (in triplicate) for a series of binders has been measured by a single lab with a single approach and has been published; (3) the set is representative of the real-world application of lead optimization: unlike in some computational benchmarks, many of the binding affinities differ by less than 1 kcal/mol and thus the precise rank ordering is challenging to reproduce; (4) it is amenable to calculations not only within the DFT/QM approach we are attempting to demonstrate, but also by widely-used classical approaches, such as FEP and molecular mechanics with generalized Born and surface area continuum solvation (MM/GBSA)[10]. It is worth noting that results using the FEP method on this system have previously been published as part of a broader validation of the method (denoted as "FEP+" in Ref. 5). As we will demonstrate, using an agnostic protocol and no parameter fitting, full DFT/QM calculations can address the problem of rank ordering with an accuracy that is superior to protocols using FEP or MM/GBSA.

**Related approaches.** As described above, full DFT/QM has typically been considered too expensive to directly apply to problems of lead optimization in protein-ligand binding. Consequently, a range of approximate DFT/QM-derived approaches have been explored such as QM/MM and other embedded QM methods[11,12], fragmentation methods such as FMO and others[13,14], semi-empirical QM including empirical tight-binding methods[15–17], and direct linear-scaling approximations[18,19]. Each technique has its strengths and there have been notable and impressive calculations with these approaches that suggest the potential of the full DFT/QM treatment[12], but overall an unambiguous improvement in accuracy with respect to classical free energy methods has yet to be convincingly shown[1]. One issue is the faithfulness of these techniques to the underlying DFT/QM, as the approximations often manifest similar weaknesses to those found with classical force fields and can further introduce other errors which can be large on the scale of discriminating small ligand-binding enthalpies. For example, both QM/MM approaches and fragmentation methods treat only a region (or multiple small regions) with explicit DFT/QM and this leads to errors arising from the boundaries, as well as in the treatment of long-range charge polarization[20]. Semi-empirical QM (SEQM), e.g., the PM6[21] and DFTB models[17], is extremely cheap and can be routinely applied to protein-ligand complexes, but reintroduces the problems of atom-specific parametrizations found in force-fields, and to date is far from being able to reliably match the accuracy of full DFT/QM, partly owing to an inability to properly describe polarization of electron density near charges. Finally, linear-scaling DFT algorithms have recently been applied to a few protein-ligand complexes[18,19]. However, while such linear-scaling approximations will eventually be required as we look towards larger and larger systems, the current numerical thresholds that control the cost introduce errors in the total energy on the scale of several kcal/mol[18], which is a similar scale to the differences in the ligand-binding enthalpies found in realistic drug interactions.

Beyond the issues annotated above, a second critical bottleneck in using these approximate DFT/QM methods in drug discovery applications is that the total wall time reported is still too long compared to the turnaround time needed for lead optimization. Since lead optimization often requires screening 100s of molecules, ideally each computation should typically not take more than an hour of wall time. However, for example, both fragment calculations (e.g. Ref. 20) and linear-scaling DFT calculations (e.g. Ref. 19) have reported wall times on the scale of a day to several days per protein-ligand complex, if a reasonable numerical accuracy is desired.

For the above reasons, full DFT/QM remains the gold standard for QM calculations, and the ultimate goal is to be able to perform such calculations for the protein-ligand complex on the time scale of an hour or less. This is what we can now achieve using the algorithms that we describe in the following section.

## Methods

**HedQM on cloud architectures.** To enable full QM calculations on protein-ligand complexes with high throughput, our strategy has been to design a core set of highly parallel DFT algorithms as part of the QSimulate-QM software platform. Rather than relying heavily on the tightly coupled nature of supercomputer hardware, we have designed our algorithms so that they can use generally available cloud computing resources, which employ commodity network interconnects. This requires optimizing the computational flow to minimize and hide communication wherever possible, as well as to ensure the calculation remains robust despite the dynamic nature of cloud computing hardware, where for example, spot instances may be taken away at any time.

Using our implementation, we can now run parallel QM using vast computing resources available on the cloud. The cloud setting carries the advantage that almost arbitrarily large computational resources can be acquired on demand by any user. For example, we have carried out calculations with over 1000+ CPU cores and over 10+ TB of distributed memory on the Amazon Web Services platform. Furthermore, no hard disk space is necessary for computation in our implementation. This makes it possible to run a full QM calculation on a realistic protein-ligand system in a fraction of an hour, and due to the robustness of our platform, we can carry out simulations even on the cheapest "spot" instances, at a cost of approximately $0.02 per core hour.

**Algorithms.** We base our full DFT/QM algorithm on the resolution-of-the-identity (RI-J) formulation[22,23] using an auxiliary density expansion[24] during orbital optimization (although we evaluate the final energy using the exact density).

We precompute the two-electron, three-index electron repulsion integrals appearing in the RI-J algorithm and keep them in distributed memory in an atom-wise sparse format. To determine the sparsity, integral bounds[25] are computed for the atomic orbital (AO) integrals with a very tight threshold ($10^{-10}$). In the largest calculations, the three-index integrals amounted to about 6TB of distributed memory in total. We perform the metric inversion in RI-J using a Cholesky decomposition for performance and numerical stability. Exchange-correlation evaluation was carried out using the LibXC library[26].

All of the one-electron matrices, including the Cholesky-decomposed metric for RI, are distributed across the MPI processes in the block-cyclic format as detailed by the ScaLapack specification[27], except for the density matrices that need to be replicated for Fock matrix evaluation. Likewise, all of the level-3 Basic Linear Algebra Subprogram (BLAS) operations

(Cholesky decomposition, matrix-matrix multiplication, etc.) are performed using the ScaLapack library as implemented in Intel MKL, except for the diagonalization step that is performed by the ELPA library[28].

To avoid convergence issues that have plagued earlier DFT calculations on very large systems[29], our algorithm uses a standard self-consistent procedure at the beginning of the calculation until the root-mean-square of the residual converges below $10^{-5}$. We then switch to an augmented Hessian algorithm with step restrictions to directly minimize the energy. In addition, we have observed that using implicit solvent in the earlier iterations stabilizes the SCF convergence in the first step; we include this using the standard conductor-like polarizable continuum (C-PCM) model using the ISWIG discretization[30–32]. The final augmented Hessian algorithm is then either run in the gas phase or with solvation as desired in the simulation. Note that the augmented Hessian algorithm is ideal for cloud hardware, especially close to convergence, because it is not communication-intensive, unlike traditional diagonalization. It also provides for reliable convergence even in the presence of the small HOMO-LUMO gaps often found in these systems, a problem previously considered highly challenging[20,29].

We have designed the inter-node communication in the DFT implementation to minimize the impact of interconnects with large latency (see below). Almost all the communication arises in the ScaLapack/ELPA libraries; some exceptions include the replication of density matrices (collective Allgather of a one-electron matrix) and Allreduce operations for small vectors.

**Runtime.** All calculations were run using r5.24xlarge instances on Amazon Web Services (AWS) unless otherwise stated. The r5.24xlarge instances are one of the memory optimized instance types and consist of two Xeon Platinum CPUs, a total of 48 physical cores and 96 threads, and 768 GB memory. The instances, instantiated in a so-called placement group, are connected by 25 Gigabit Ethernet and used collectively by our message passing interface (MPI)-parallelized program. In addition to inter-node parallelization, the program is threaded for intra-node concurrency. We used 4 MPI processes per instance with up to 24 threads dedicated to each MPI process. The choice of instances is primarily based on the large memory requirements for the calculations and on the fact that there is no need for local hard disk storage (see above). We used 12 r5.24xlarge spot instances for each of the MCL1 protein-ligand calculations, which took about 40 minutes on average; the cost for each calculation for the full complex was around US$10.

**Preparation and scoring workflow.** Starting from the PDB structure of MCL1 [PDB ID: 4HW3], the subsequent computational procedure required several steps: (1) protein preparation; (2) ligand docking; (3) refinement of the protein-ligand complex (i.e., geometry optimization); (4) scoring of the protein-ligand binding affinities. Additional details are provided in the supplementary information, and we highlight here only the essential points. To prepare the protein structure, (1) we neutralized residues that were far (> 7.5 Å) from the ligand, to better reflect their effective charges when placed in a solvated environment with counterions. For ligand docking (2) we used Smina[33] to generate 10 poses per ligand (220 poses in total). To refine the protein-ligand complex (3) we used an Amgen in-house MM/GBSA protocol that was interfaced with Amber18[34]. Finally, for scoring binding in the protein-ligand complex (4), we used three different approaches for comparison: MM/GBSA[10], the semi-empirical method GFN1-xTB using the implementation provided by Grimme's group[16], and DFT/QM. In each case, the binding energy was obtained by subtracting the energy of the ligand from the energy of the protein-ligand complex (note that since there was only a single protein, its energy did not need to be computed, and we did not do so in DFT/QM or SEQM).

**Initial scoring procedure.** All 220 Smina poses were initially scored using MM/GBSA. Ligand strain was reduced by optimizing the bound ligand geometry. Out of these initial 220 poses, we retained only those within 7.5 kcal/mol of the lowest energy MM/GBSA pose by ligand, reducing the set to 67 poses. These 67 poses were then scored using either DFT/QM or SEQM to obtain the best pose for each of the 22 ligands and the final relative ligand binding energy.

**DFT/QM.** DFT calculations on ligands bound to MCL1 were carried out using the revised PBE (revPBE) functional[35] together with Grimme's dispersion correction [D3(BJ)][36]. We used revPBE-D3(BJ) because it has been shown to perform well among pure DFT functionals in systems with non-covalent interactions[37,38]. The def2-SVP basis set[39] was used for atoms belonging to residues within 10 Å of the ligand, and a minimum basis set derived from cc-pVTZ (MINAO) was used for the rest of the atoms[40]. Increasing the size of the region that included the larger def2-SVP basis did not change the results (see supplementary information). Using this basis, the numbers of primary and auxiliary Gaussian basis functions were approximately 15,200 and 49,000, respectively. The number of electrons and atoms were respectively approximately 9,400 and 2,470. We also carried out the DFT calculations with and without solvation using C-PCM and the ISWIG discretization.

## Results

**Ranking ligand binding to MCL1.** Ligand binding affinities to MCL1 (induced myeloid leukemia cell differentiation protein) have previously been used as one of the validation systems for Schrodinger's FEP+ protocol. Ref. 41 provides a set of 42 ligands together with a measured set of dissociation constants ($K_i$) based on the skeleton shown at the bottom right of Fig. 1. To restrict the computational costs to our computational budget, out of the 42 ligands reported in their paper, we used the set of 22 homologous ligands where the aryl group (Ar) contains a single ring. These 22 ligands, along with associated $K_i$ values, are shown in Fig. 1. The ligands vary in size from 38 to 47 atoms. Note that the chosen ligands here span a 500-fold range in $K_i$, which is realistic for a later stage drug discovery effort. Distinguishing between these ligands and obtaining the correct rank ordering is thus a relevant test of the computational methodology. We characterize the performance of DFT/QM and its potential for ligand screening by computing the ligand binding affinities using a single structure, single energy application and use the results to rank order the ligands. A graphic of the entire system simulated using DFT/QM is shown in Fig. 2.

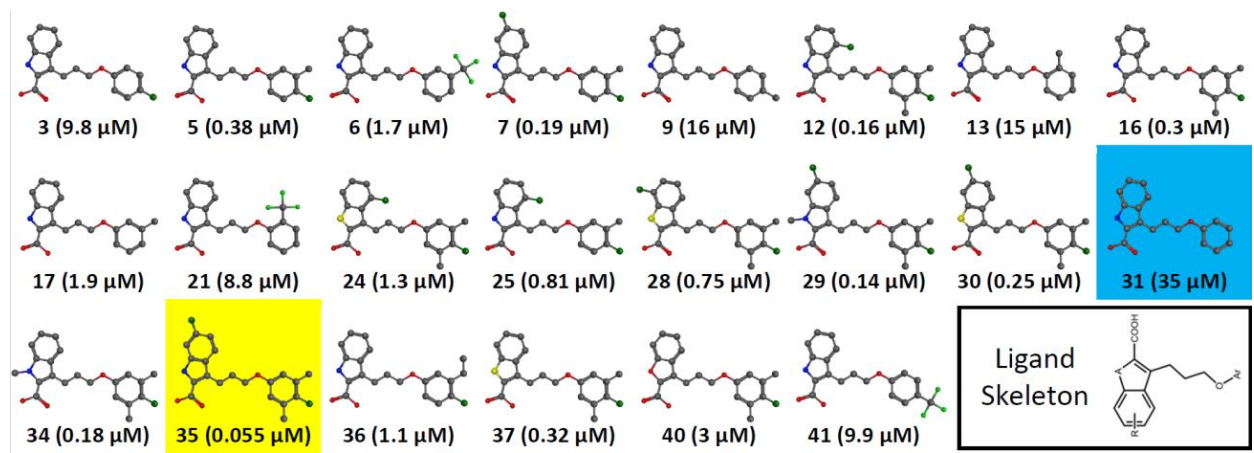

**Figure 1.** Set of 22 ligands together with their K$_i$ values. The numeric labels of the ligands are the ones taken from Ref. 41. The strongest and weakest binders are highlighted, as measured experimentally. The ligand skeleton used in Ref. 41 is shown at the bottom right. In this work, we considered the subset of 22 ligands where the Ar group contained a single ring.

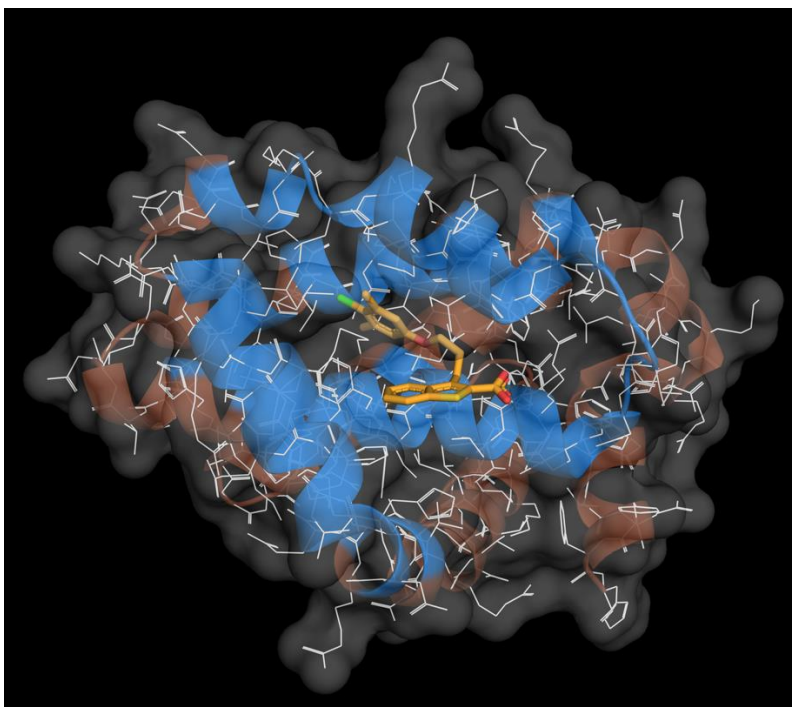

**Figure 2.** The MCL1 binding domain (10 Å cutoff shown with blue ribbons) with a representative ligand (#37 from Fig. 2) in the binding pocket. The full ligand plus domain incorporates approximately 2500 atoms.

As a comparison, we will compare to the FEP results from Ref. 5 (using the OPLS2.1 force field) for rank ordering these ligands. However, on a theoretical level, the FEP results do not strictly provide an apples-to-apples comparison to our DFT/QM approach because they compute the free energy, rather than the electronic energy. Thus, for additional insight, we also present the binding energies within the single structure approach using the MM/GBSA method (the potential energy calculated using the force field augmented by a generalized Born treatment of the solvation energy), and a semi-empirical QM method, GFN1-xTB, that is designed specifically to improve the description of systems with non-covalent interactions[16]. One additional variable in the single structure approach is the choice of the unbound ligand geometry. We studied three choices with QM: (A) the ligand fixed at the geometry of the bound complex; (B) geometry of the free ligand optimized starting from the geometry of the bound complex to the nearest local minimum; and (C) geometry of the free ligand optimized to the global lowest energy conformer. Note that all optimizations were performed using MM/GBSA, the same method used to create the starting protein-ligand complex geometry, to maintain consistency.

In Fig. 3 we show the rank orderings obtained for the 22 ligands using FEP, MM/GBSA, SEQM (GFN1-xTB), and QM (revPBE-D3(BJ)) together with three metrics that summarize the correlations: the coefficient of determination ($R^2$), the Spearman rank correlation coefficient ($\rho$), and the predictive index (PI). The FEP results are taken from Ref. 5. As a baseline, simple docking using the Smina docking package results in an $R^2$ close to 0, when scored using the default Autodock Vina scoring function[42], while docking and scoring using the Glide program[43] using the SP scoring function results in an $R^2$ around 0.42.

The Spearman's $\rho$ coefficient summarizes the quality of the rank orderings of the methods. A correlation of 1 implies that the computed and experimental orderings of the binding affinities are identical. In fact, all methods perform reasonably well for the rank orderings, but the rank ordering obtained using the DFT/QM is appreciably better (10%) than that obtained using the other methods. FEP and MM/GBSA perform almost identically suggesting that entropic effects are either small or well correlated with enthalpic effects in this system, and SEQM is slightly better than both.

Another metric with similar purpose, the Predictive Index (PI)[44], also reflects rank ordering, and varies from 1 (perfect ranking) to 0 (entirely random), just like Spearman. But the Predictive Index weights contributions to the index by the experimental distance between the points being compared, thereby, e.g. reducing the importance of rank ordering for points that are effectively the same within experimental error. Surprisingly, evaluated using PI, FEP is inferior to all three other approaches, even MM-GBSA.

The $R^2$ value is a widely-reported measure of the correlation between the experimental and computational predictions. Although $R^2$ is commonly presented, it is often a poor metric for characteristics of real importance, such as rank ordering. Again, by this metric, we find that DFT/QM performs significantly better than any other method; FEP+ and MM/GBSA perform almost identically to each other; and SEQM is slightly better than those methods, but still inferior to DFT.

Note that while the predictions using methods other than FEP are quite capable of correctly predicting rank ordering, only FEP is expected (in cases where it works well) to fit a line with a slope of around 1 and an intercept around (0,0). The magnitudes of predictions using the other methods result in significantly different slopes and intercepts, as has been seen previously[1,20]. However, in practical use for later stage drug discovery, reliable rank ordering is appreciably more important than metrics like mean error.

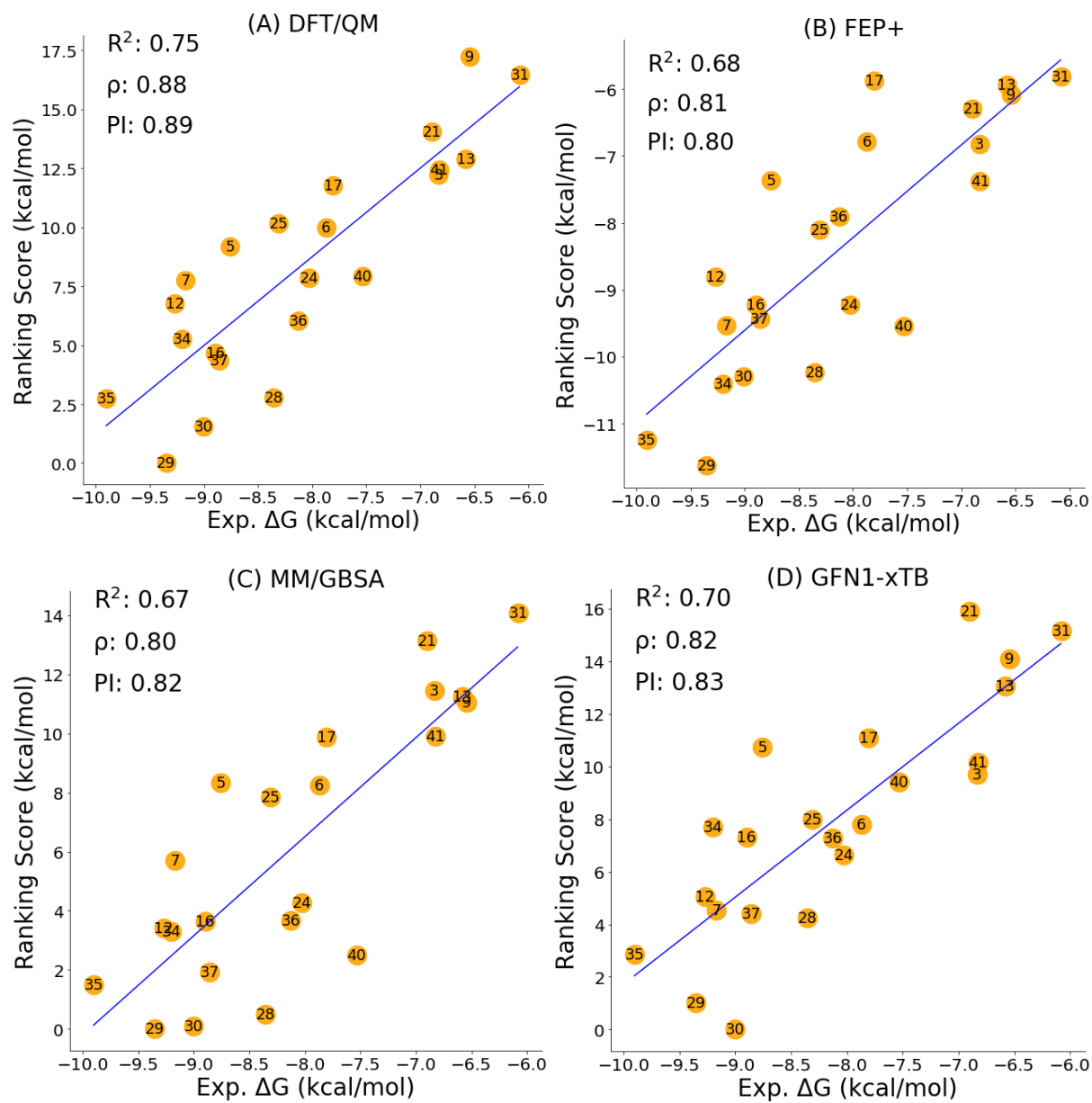

**Figure 3.** Predictions using various methods for a set of 22 ligands binding to MCL1. (A) DFT/QM, (B) FEP+ (from Ref. 5), (C) MM/GBSA, and (D) GFN1-xTB. The numbers in the circles correspond to the ligand numbers in Figure 1. DFT/QM provides the best correlations out of all the methods employed. For DFT/QM and GFN1-xTB, ligand strain is computed relative to the global minimum conformer, and ligand desolvation is taken into account. The predictive index (PI) reflects the ability to rank order the data, with each pairwise contribution weighted by the experimental distance between the corresponding pair of experimental observations[44].

We can obtain some insight into the physics involved in the DFT/QM ligand binding scoring by studying the component contributions to the energy. For example, we can examine the effect of solvation and ligand geometry relaxation (both to local and global minima) on the DFT rank orderings and $R^2$ values. The correct computation of the ligand-binding enthalpy should include both desolvation and the (global) ligand strain effects. In Table 1, we show the $R^2$ and Spearman rank coefficient obtained in the DFT/QM calculation as a function of the inclusion of these two contributions. For comparison, we also show the effect on the SEQM method GFN1-xTB. As we observe, removing either the ligand relaxation or solvation contribution worsens the $R^2$ and Spearman coefficients. This suggests that the good performance of DFT/QM is correlated with the correct physics. The best $R^2$ and $\rho$ (0.75 and 0.88) are obtained when solvation is accounted for and the global minimum for each ligand is used to incorporate strain. In the case of GFN1-xTB, although correlations comparable to the classical methods are reached, the results do not seem highly correlated with the contributions that are included.

**Table 1.** $R^2$ and Spearman rank correlation $\rho$ for 6 different protocols to account for ligand strain and desolvation using DFT/QM and GFN1-xTB. Vacuum = no solvation; water = with solvation; b = ligand geometry is bound geometry; o = ligand geometry is local minimum near bound geometry; c = ligand geometry is lowest-energy conformer. DFT/QM yields the highest $R^2$ and rank correlation when both solvation and full ligand relaxation is included, consistent with what would be expected from physical arguments.

| solvation | ligand | GFN1-xTB $R^2$ | DFT/QM $R^2$ | GFN1-xTB $\rho$ | DFT/QM $\rho$ |
|---|---|---|---|---|---|
| vacuum | b | 0.67 | 0.71 | 0.78 | 0.84 |
| vacuum | o | 0.69 | 0.71 | 0.79 | 0.83 |
| vacuum | c | 0.69 | 0.60 | 0.81 | 0.78 |
| water | b | 0.69 | 0.71 | 0.84 | 0.84 |
| water | o | 0.66 | 0.72 | 0.82 | 0.85 |
| water | c | 0.70 | 0.75 | 0.82 | 0.88 |

As a further test to support the hypothesis that the DFT/QM approach is obtaining good correlations for the right reasons, we assessed the correctness of the top pose (the one predicted to be energetically most favorable) for all 22 ligands across 12 different methodologies. The results are summarized in Fig. 4. While co-crystallized structures for all ligands are not available, the homology between the ligands suggests that we can assess the appropriateness of poses based on their similarity to the available co-crystallized ligand pose, and in particular by verifying that the fused 5/6 ring, linker, and aryl group are aligned with the crystal structure ligand. The red boxes indicate a protocol/ligand combination that results in an "incorrect" (i.e., dissimilar) pose being selected as the most energetically favorable one. The protocols differ with respect to method (QM vs. SEQM), the inclusion of solvation (vacuum vs. water), and ligand strain effects (b = ligand geometry is bound geometry; o = ligand geometry is local minimum near bound geometry; c = ligand geometry is lowest-energy conformer). Some examples of incorrect poses are shown in Fig. 5. From the 6 protocols considered in DFT/QM, only 3 manage to pick out the correct pose for all 22 ligands, and out of those, only two give reasonable $R^2$ values. The ligand for which all GFN1-xTB protocols are unable to pick out the correct pose is the rightmost image in Fig. 5. It is reassuring that the best results are obtained when DFT is applied in the most physically sensible manner (accounting for solvation and ligand strain relative to the global minimum). SEQM (GFN1-xTB), on the other hand, fails to pick the correct pose for all ligands with any of the protocols.

| | | | | DFT/QM | | | | | | GFN1-xTB/SEQM | | | | | |
|---|---|---|---|---|---|---|---|---|---|---|---|---|---|---|---|
| | | | | vacuum | | | water | | | vacuum | | | water | | |
| Lig. # | kcal/mol | # poses | $K_i$ | b | o | c | b | o | c | b | o | c | b | o | c |
| 3 | -6.83 | 4 | 9.80 | | | | | | | | | | | | |
| 5 | -8.76 | 4 | 0.38 | x | | | x | | | x | x | | x | x | |
| 6 | -7.87 | 4 | 1.70 | x | | | x | | | | | | | | |
| 7 | -9.17 | 4 | 0.19 | | | | | | | | | | | | |
| 9 | -6.54 | 2 | 16.00 | | | | | | | | | | | | |
| 12 | -9.27 | 2 | 0.16 | | | | | | | | | | | | |
| 13 | -6.58 | 2 | 15.00 | | | | | | | | | | | | |
| 16 | -8.90 | 2 | 0.30 | | | | | | | | | | | | |
| 17 | -7.81 | 3 | 1.90 | x | | | x | | | | | | | | |
| 21 | -6.90 | 5 | 8.80 | | | | | | | | x | | | | |
| 24 | -8.03 | 3 | 1.30 | | | | | | | | | | | | |
| 25 | -8.31 | 2 | 0.81 | x | x | | | | | x | x | | | | |
| 28 | -8.36 | 4 | 0.75 | | | | | | | x | x | x | x | x | x |
| 29 | -9.35 | 3 | 0.14 | | | | | | | | | | | | |
| 30 | -9.01 | 2 | 0.25 | | | | | | | | | | | | |
| 31 | -6.08 | 3 | 35.00 | | | | | | | | | | | | |
| 34 | -9.20 | 3 | 0.18 | | | | | | | | | | | | |
| 35 | -9.90 | 2 | 0.06 | | | | | | | | | | | | |
| 36 | -8.13 | 3 | 1.10 | | | | | | | | | | | | |
| 37 | -8.86 | 4 | 0.32 | | | | | | | | | | | | |
| 40 | -7.53 | 2 | 3.00 | | | | | | | | | | | | |
| 41 | -6.83 | 4 | 9.90 | | | | | | | | | | | | |

**Figure 4.** Pose identification for the 22 ligands (involving 67 total poses) using 6 different solvation protocols and ligand geometries (see main text and Figure 3 for description of protocols) for DFT/QM and GFN1-xTB. Green indicates that the "correct" pose is selected (as judged by similarity to co-crystallized ligand pose), while red indicates an "incorrect" pose. DFT/QM finds the correct pose for all 22 ligands when the relevant physical contributions of ligand solvation and strain relative to the lowest-energy conformer are included (protocol marked in yellow).

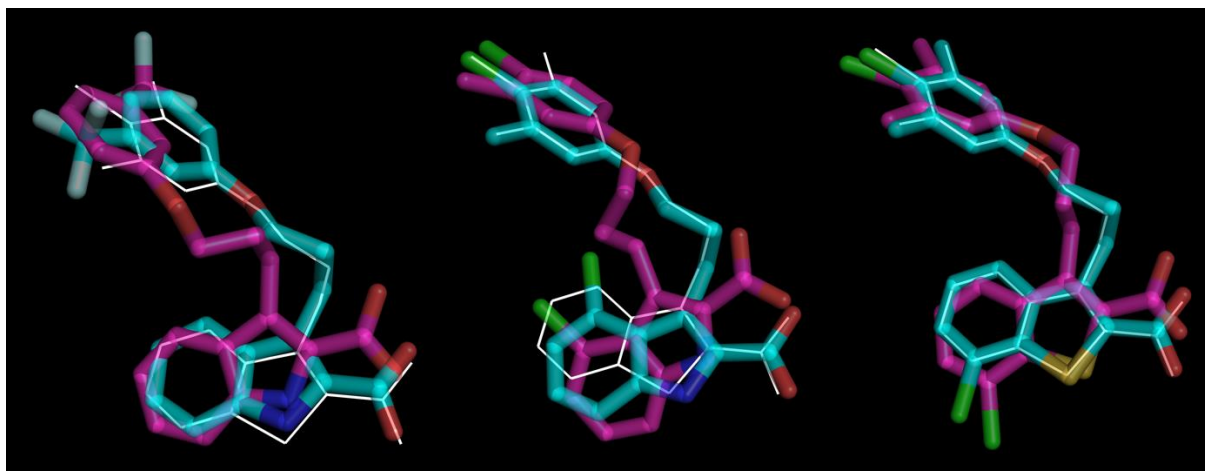

**Figure 5.** Examples of 'incorrect' poses (in magenta) overlaid with 'correct' poses (in cyan) for 3 of the 22 ligands considered in this study: 6, 25, 28 (from left to right). The crystal structure ligand from PDB ID: 4HW3 is shown with white lines.

## Discussion

It is undeniable that quantum mechanics—as embodied, for example, in modern DFT/QM calculations—is, in principle, the best approach to exploring many chemical phenomena. However, the enormous resource requirements of such calculations and the associated low throughput have traditionally limited the usefulness of QM for drug discovery. The approach we have described herein greatly ameliorates the throughput issue through the harnessing of large scale, distributed memory cloud resources. This means that we can examine a significant number of protein-ligand interactions using DFT/QM, on the timescale of a one-day project. While each protein-ligand calculation described in this work cost approximately $10 and took approximately 40 minutes to complete, the turnaround time could be decreased further (at larger dollar cost) simply by increasing the number of cloud instances. Highly efficient distributed-memory QM (hedQM) on full protein-ligand systems is thus now practical.

Because our approach does not require the use of additional approximations to the DFT/QM, we can also now understand the influence of quantum mechanics on protein-ligand binding free from other theoretical and numerical issues. Application of hedQM in an agnostic fashion to a series of inhibitors of the MCL1 protein and comparison to FEP results obtained in a previous publication for the same series of ligands demonstrates that DFT/QM performs better for this system by relevant metrics, such as $R^2$, Spearman rank, or PI.

Although our work suggests that hedQM calculations can now compete with FEP in terms of computational cost, and further carry the promise to substantially expand the domain of applicability versus FEP, it is important to note that with hedQM we are still practically limited to calculating potential energies for a relatively small number of poses or conformers. As a result, we cannot currently directly compute the configurational entropy for the system, and we need to rely on a combination of cancellation of contributions when comparing multiple ligands binding to a single receptor protein and corrections to the DFT/QM enthalpy. The good results we obtain using hedQM versus FEP in this communication demonstrate that the approach we describe is reasonable, however, because it is dependent on the geometry of a single (or a small number of) structures, in future work, we will further characterize this protocol across a broader set of proteins and ligands to evaluate its robustness in larger-scale protein-ligand screening. In addition, the combination of hedQM with approaches to obtain quantum mechanically corrected free energy contributions[1], as well as machine learning approaches to reduce the number of fully quantum mechanical calculations[45,46], are also natural avenues to explore, to extend this work to a more rigorous estimation of entropic contributions. Importantly, our work now places such investigations for drug discovery firmly within the domain of practical possibility, rather than simply fantasy.

## Acknowledgments

The authors thank Eric Kessler, Chris Downing, David Kanter, and other members of the Amazon Web Services and Amazon Quantum Solution Lab for technical support and helpful conversations. We would also like to thank Yax Sun, Kai Zhu, Huan Rui, Lei Jia, and David Khachatrian at AMGEN Research for helpful comments.

## Competing interests

The authors declare the following competing interests: Y.W., D.A.P., and T.S. are employees of Quantum Simulation Technologies, Inc. (QSimulate), in which G.K.-L.C. and T.S. are significant shareholders. G.K.-L.C. is a consultant for QSimulate. N.M. declares no competing interest.

## Author contributions

N.M., G.K.-L.C., T.S. designed the study, N.M., Y.W., T.S. performed the calculations, all authors participated in the analysis of the results, all authors participated in the writing of the manuscript.